\documentclass[12pt]{article}
\usepackage[dvips]{graphicx}
\usepackage{amssymb}
\usepackage{amsmath}

\begin{document}

\begin{center}
\textbf{\Large Properties of the quasistationary universe in
context of the Big-bang cosmology problems}
\end{center}

\vspace*{0.2cm}

\begin{center}

{\large V.E. Kuzmichev and V.V. Kuzmichev}\\[0.3cm]

Bogolyubov Institute for Theoretical Physics, National Academy of
Sciences of Ukraine, Kiev, 03143 Ukraine

\end{center}

\vspace*{0.2cm}

\begin{abstract}
Old and new puzzles of cosmology are reexamined from the point of
view of quantum theory of the universe developed here. It is shown
that in proposed approach the difficulties of the standard
cosmology do not arise. The theory predicts the observed
dimensions of the nonhomogeneities of matter density and the
amplitude of the fluctuations of the cosmic background radiation
temperature in the Universe and points to a new quantum mechanism
of their origin. It allows to obtain the value of the deceleration
parameter which is in good agreement with the recent SNe Ia
measurements. The theory explains the large value of entropy of
the Universe and describes other parameters.
\end{abstract}

\vspace*{0.5cm}

\section{Introduction}
The classical cosmology based on the equations of general
relativity involving the principles of thermodynamics,
hydrodynamics, the plasma theory and the field theory comes across
a number of conceptual difficulties known as the problems of
standard Big-bang cosmology \cite{1,2,3,4}. These are the problems
of singularity, size, age, flatness, total entropy and total mass
of the Universe, large-scale structure, dark matter, isotropy of
the cosmic background radiation and others. The various models
were proposed for the solution of these problems. The inflationary
model \cite{1,2} is the most popular one. There are alternative
approaches which use the idea that in the early Universe the
fundamental constants (velocity of light, gravitational constant,
fine-structure constant) had the values different from the modern
\cite{5,6}.

The observations of type Ia supernovae (SNe Ia) indicate that our
Universe is accelerating \cite{P,R}. This conclusion which
appeared as partly unexpected for the cosmologists a few years ago
nowadays practically does not called in question \cite{T}. The
concept of a dark energy was proposed for the explanation of this
phenomenon \cite{O,B} and the modern investigations in this field
are directed toward filling of this idea with concrete contents
\cite{C,BB,NW}.

The presence of the cosmological problems points to incompleteness
of our knowledge about the Universe. It is generally accepted that
the conclusions of classical theory of gravity cannot be
extrapolated to the very early epoch. At the Planck scales one
must take into account the quantum effects of both matter and
gravitational fields.

There cannot be any doubt  that our Universe today contains the
structural elements which bear the traces of comprehensive quantum
processes in preceding epochs. The small cosmic microwave
background anisotropy and observed large-scale structure of the
Universe \cite{4} can be given as necessary examples (see below).

The application of basic ideas underlying quantum theory to a
system of gravitational and matter fields runs into difficulties
of a fundamental character which do not depend on the choice of a
specific model. The problem of separation of true degrees of
freedom under the construction of quantum gravity becomes of
fundamental importance \cite{7}. It is commonly thought that the
main reason behind such difficulties is that there is no natural
way to define a spacetime event in general covariant theories
\cite{8}. At the present time these difficulties are not overcome
in the most advanced versions of quantum gravity. Also the quantum
gravity cannot rely on experimental data \cite{9}. Therefore it is
appropriate to construct the consistent quantum theory within the
framework of the simple (toy) exactly soluble cosmological model.
As it is well known the model of a homogeneous, isotropic universe
(Friedmann-Robertson-Walker model) describes good enough the
general properties of our Universe. In this paper we study the
model of quantum universe proposed in \cite{10,11,12,12b}. It does
not meet with the problems mentioned above and comes to the FRW
model with positive spatial curvature within the limits of large
quantum numbers.

In Section 2 we propose the method of removing ambiguities in
specifying the time variable in the FRW model by means of
modification of the action functional and find the solutions of
obtained classical field equations. The Section 3 is devoted to
the quantum theory for a system of gravitational and matter
fields. Here we formulate the equation which is an analog of the
Schr\"{o}dinger equation and turns into the Wheeler-DeWitt one for
the minisuperspace model in special case. We concentrate our
attention on study of the quantum universe which can be found in a
region that is accessible to a classical motion inside the
effective barrier formed by the interaction of the fields. We
discuss the properties of the wave function of the universe and
study the universe in low-lying and highly exited quasistationary
states on the basis of exact solution of proposed quantum
equation. In this section we calculate the proper dimension of the
nonhomogeneities of matter density and the amplitude of the
fluctuations of the cosmic background radiation temperature in
highly exited state of the universe and propose a new possible
quantum mechanism of their origin. The results are compared with
the observed parameters of our Universe. The flatness of the
Universe and the large value of the entropy today receive their
natural explanation. The observed accelerated expansion emerges as
a macroscopic manifestation of the quantum nature of the Universe.

Throughout the paper the notation \textit{Universe} (with capital
letter U) relates to our Universe, while \textit{universe} (with
small letter u) corresponds to arbitrary cosmological system of
considered type.

\section{Classical description}

\subsection{Coordinate condition and basic equations}

For simplicity we restrict our study to the case of minimal
coupling between geometry and the matter. Considering that scalar
fields play a fundamental role both in quantum field theory and in
the cosmology of the early Universe we assume that, originally,
the Universe was filled with matter in the form of a scalar field
$\phi$ with some potential $V(\phi)$. As we shall see the
replacement of the entire set of actually existing massive fields
by some averaged massive scalar field seems physically justified.
Assuming that the field $\phi$ is uniform and the geometry is
defined by the Robertson-Walker metric, we represent the action
functional in the conventional form
\begin{eqnarray}
  S  = \int d\eta \,\left[\pi _{a}\,\partial_{\eta}a + \pi _{\phi }\,
  \partial_{\eta}\phi  -  \mbox{H} \right].
\label{1}
\end{eqnarray}
Here $\eta$ is the time parameter that is related to the
synchronous proper time $t$ by the differential equation $dt =
N\,a\,d\eta$, where $N(\eta)$ is a function that specifies the
time-reference scale, $a(\eta)$ is a scale factor; $\pi _{a}$ and
$\pi _{\phi}$ are the momenta canonically conjugate with the
variables $a$ and $\phi$, respectively. The Hamiltonian $\mbox{H}$
is given by
\begin{equation}
    \mbox{H} = \frac{1}{2}\,N\,\left[ -\,\pi _{a}^{2}
       + \frac{2}{a^{2}}\,\pi _{\phi }^{2} - a^{2}
       + a^{4}\,V(\phi ) \right] \equiv N\, {\cal R},
\label{2}
\end{equation}
where the $a$ is taken in units of the length $l = \sqrt{2 /
3\,\pi }\, l_{Pl}$, $l_{Pl}$ is the Planck length, and $\phi$ in
units of $\tilde \phi = \sqrt{3 / 8\,\pi \,G}$. The energy density
will be measured in units $(\tilde{\phi}/l)^{2} =
(9/16)\,m_{Pl}^{4}$.

The function $N$ plays the role of a Lagrange multiplier, and the
variation $\delta S/\delta N$ leads to the constraint equation
${\cal R} = 0$. The structure of the constraint is such that true
dynamical degrees of freedom cannot be singled out explicitly. In
the model being considered, this difficulty is reflected in that
the choice of the time variable is ambiguous (the problem of
time). For the choice of the time coordinate to be unambiguous,
the model must be supplemented with a coordinate condition. When
the coordinate condition is added to the field equations, their
solution can be found for chosen time variable. However, this
method of removing ambiguities in specifying the time variable
does not solve the problem of a quantum description. Therefore we
shall use another approach and remove the above ambiguity  with
the aid of a coordinate condition imposed prior to varying the
action functional. We will choose the coordinate condition in the
form
\begin{equation}
  g^{00}\left(\partial_{\eta}T\right)^{2} = \frac{1}{a^{2}}\,, \quad
  \mbox{or} \quad \partial_{\eta}T = N,
  \label{3}
\end{equation}
where $T$ is the privileged time coordinate, and include it in the
action functional with the aid of a Lagrange multiplier $P$
\begin{equation}
  S = \int \! d\eta\, \left[\,\pi _{a}\,\partial_{\eta}a + \pi _{\phi }\,
  \partial_{\eta}\phi + P\,\partial_{\eta}T - {\cal H}\,\right],
\label{4}
\end{equation}
where
\begin{equation}
 {\cal H} = N\,[\,P + {\cal R}\,]
\label{5}
\end{equation}
is the new Hamiltonian. The constraint equation reduces to the
form
\begin{equation}
  P + {\cal R} = 0.
  \label{51}
\end{equation}
Parameter $T$ can be used as an independent variable for the
description of the evolution of the universe. Corresponding
canonical equations reduce to the form
\begin{eqnarray}
         \partial_{T} a & = & -\,\pi _{a}, \qquad \
         \partial_{T} \pi _{a}  =  \frac{2}{a^{3}}\,\pi ^{2}_{\phi } + a -
             2\,a^{3} V (\phi), \nonumber\\
     \partial_{T} \phi  & = & \frac{2}{a^{2}}\,\pi _{\phi }, \qquad
     \partial_{T} \pi _{\phi } = -\,\frac{a^{4}}{2}\,\frac{dV(\phi)}{d\phi },
     \nonumber\\
         \partial_{T} T & = & 1, \qquad \qquad \partial_{T} P = 0.
\label{52}
\end{eqnarray}
Integrating the equation for $P$, we obtain $P = E / 2$, where $E$
is a constant and the multiplier $1/2$ is introduced for further
convenience. The full set of equations for the model in question
becomes \cite{10,11}
\begin{equation}
  \left(\partial_{T} a \right)^{2} - \frac{a^{2}}{2}\,\left(\partial_{T} \phi
  \right)^{2} + U = E,
\label{6}
\end{equation}
\begin{equation}
  \partial^{2}_{T} \phi  + \frac{2}{a}\,\left(\partial_{T} a \right)
  \left(\partial_{T} \phi \right) +
   a^{2}\,\frac{dV}{d\phi } = 0,
\label{7}
\end{equation}
where $U = a^{2} - a^{4}\,V(\phi )$. Equation (\ref{6}) represents
the Einstein equation for the $\left(^{0}_{0}\right)$ component,
while equation (\ref{7}) is the equation of motion
$T^{\mu}_{0;\mu} = 0$ for the field $\phi $. ($T^{\mu}_{\nu}$ is
the energy-momentum tensor of the scalar field.)

From the analysis of the Einstein equations for this model it
follows that inclusion of the coordinate condition (\ref{3}) in
the action functional leads to the origin of the additional
energy-momentum tensor in these equations
\begin{equation}
  \tilde T^{0}_{0} = \frac{E}{a^{4}},\quad
 \tilde T^{1}_{1} = \tilde T^{2}_{2} = \tilde T^{3}_{3} =
 -\,\frac{E}{3\, a^{4}},\quad \tilde T^{\mu }_{\nu } = 0
  \ \ \mbox{for} \ \  \mu \neq \nu,
  \label{8}
\end{equation}
that can be interpreted as the energy-momentum tensor of
radiation. In the ordinary units $E$ is measured in $\hbar$. The
choice of radiation as the matter reference frame is natural for
the case in which relativistic matter (electromagnetic radiation,
neutrino radiation, etc.) is dominant at the early stage of
Universe evolution. If our Universe were described by the model
specified by action functional (\ref{4}), it would be possible to
relate the above radiation at the present era to cosmic microwave
background.

\subsection{Solutions}
A feature peculiar to the model in question is that it involves a
barrier in the variable $a$ described by the function $U$. This
barrier is formed by the interaction of the scalar and
gravitational fields. It exists for any form of the positive
definite scalar-field potential $V(\phi )$ and becomes
impenetrable on the side of small $a$ in the limit $V \rightarrow
0$. In general case ($E \neq 0$) there are two regions accessible
to a classical motion: inside the barrier ($a \leq a_{1}$) and
outside the barrier ($a \geq a_{2}$), where $a_{1}$ and  $a_{2}$
are the turning points ($a_{1} < a_{2}$) specified by the
condition $U = E$. The set of equations (\ref{6}) and (\ref{7})
determines the $a$ and $\phi$ as the functions of time $T$ at
given $V(\phi)$. When the rate at which scalar field changes is
much smaller than the rate of universe evolution, i.e.
$(\partial_{t} \phi)^{2} \ll 2\,H^{2}$, where $H = \partial_{t}a /
a$ is the Hubble constant, and $|\partial_{t}^{2}\phi| \ll |dV /
d\phi|$, the equations (\ref{6}) and (\ref{7}) become
\begin{equation}
  \left(\partial_{T} a \right)^{2} + U = \epsilon,
  \label{9}
\end{equation}
\begin{equation}
  \frac{3}{a}\,H\,\partial_{T} \phi = - \frac{dV}{d \phi},
  \label{91}
\end{equation}
where $\epsilon$ and $U$ depend parametrically on $\phi$. In the
zero-order approximation $\epsilon  = E$. The solution to equation
(\ref{6}) can be refined by taking into account a slow variation
of the field $\phi $ with the aid of the equation
\begin{equation}
    -\,\frac{a^{2}}{2}\,\left(\partial_{T} \phi \right)^{2} + \epsilon (\phi) = E,
\label{10}
\end{equation}
where $\epsilon$ stands for a potential term.

The solutions of the equation (\ref{91}) which determine the
scalar field dynamics were studied in the inflationary models
\cite{2,4}. The solution of the equation (\ref{9}) at fixed value
of $\phi$ can be represented in the form
\begin{equation}
a(t) = \left[\frac{1}{2 V} + \frac{y}{4 V}\ \exp \left\{2\,
\sqrt{V}(t - t_{in})\right\} + \frac{1 - 4 V \epsilon }{4 V y}\
\exp \left\{-\,2\, \sqrt{V}(t - t_{in})\right\}\right]^{1/2},
\label{92}
\end{equation}
where we denote
\begin{equation}
  y = 2\,\sqrt{V ( \epsilon -
\alpha^{2} + \alpha^{4}\,V)} + 2 V \alpha^{2} - 1.
  \label{93}
\end{equation}
Here $\alpha = a(t_{in})$ gives the initial condition for some
instant of time $t = t_{in}$. At $a(0) = 0$ and $a(t_{in}) =
a_{2}$ the corresponding scale factors are given in \cite{10,11}.
The solution (\ref{92}) shows that in the region $a > a_{2}$ the
universe expands in the de Sitter mode from the point $a = a_{2}$,
but in the region $a < a_{1}$ it evolves as
    $ a(t) \simeq \left[2\,\sqrt{\epsilon }\,t \right]^{1/2}$ for
    $2\,\sqrt{V}\,t \ll 1,$
that describes the evolution of the universe which density was
dominated by radiation and as $a(t) = a_{1} - \zeta(t)$ with
$\zeta(t) \sim t^{2}$ near the point of maximal expansion $a =
a_{1}$. The estimations for $a_{1}$ demonstrate that at small
enough $V$ the value $a \sim a_{1}$ can reach the modern values of
the scale factor in our Universe. So, for the state of the
universe with $\epsilon \sim 1 / 4V$ and $V \sim 10^{-5}$
GeV/cm$^{3} = 6.1 \times 10^{-123}$ (the mean matter-energy
density in our Universe at the present era) we have $a_{1} \sim
\sqrt{1/2V} \sim 10^{61} \sim 10^{28}$ cm.

In the extreme case of $E = 0$, where there is no radiation, the
region $a \leq a_{1}$ contracts to the point $a = 0$, and the
expansion can proceed only from the point $a = a_{2}$ and  the
region $a < a_{2}$ cannot be treated in terms of classical theory.
Such models were widely enough studied by many authors (see, for
example, \cite{2,3,13,14}).

We concentrate our attention on study of the properties of the
universe which is characterized by the nonzero values of $E$ (and
$\epsilon$) at the initial instant of time and can be found in a
region that is accessible to a classical motion inside the
barrier.

The evolution of the universe depends on the initial distribution
of the classical field $\phi$ and its subsequent behavior as a
function of time. The solutions of the equation (\ref{91}) for $V
\sim \phi^{n}$ give evidence that the $\phi$ decreases with time
\cite{2,3}. From equations (\ref{7}) and (\ref{10}), it follows
that the inequality $\partial_{T} V +
\partial_{T} \epsilon / a^{4} < 0$ holds in the expanding
universe. If $V$ decreases with time, $\epsilon$ can increase. Let
us estimate $\epsilon $ by using the relation $\epsilon \simeq
\tilde T^{0}_{0} a^{4}$. In our Universe, with $a \sim 10^{28}$
cm, the main contribution to the radiation-energy density comes
from cosmic microwave background with energy density $\rho
_{\gamma }^{0} \sim 10^{- 10}$ GeV/$\mbox{cm}^{3}$. Setting
$\tilde T^{0}_{0} = \rho _{\gamma}^{0}$, we find that, at the
present era, the result is $\epsilon = \epsilon _{\gamma } \sim
10^{117}\,\hbar$. In the early Universe with $a \sim 10^{- 33}$ cm
and the Planck energy density we have $\epsilon \sim \hbar$. It
indicates that $\epsilon $ should increase in the evolution
process. This increase can be explained by a considerable
redistribution of energy between the scalar field and radiation at
the initial stage of Universe existence. Quantum theory is able to
account for this phenomenon in a natural way as spontaneous
transition from one quantum state of the universe to another (see
below).

Let us estimate the size of the classical universe in the region
$a \leq a_{1}$ taking into account that the parameter $\epsilon$
can take different values. Such universe by definition cannot
tunnel through the barrier into the region $a \geq a_{2}$ and by
this reason will evolve in time remaining inside the barrier.
Assuming for simplicity that the universe has expanded according
to the law  $ a(t) = \left[2\,\sqrt{\epsilon }\,t \right]^{1/2}$
for two instants of time $t_{0}$ and $t_{p}$ we obtain
\begin{equation}
  \frac{a(t_{0})}{a(t_{p})} =
  \left(\sqrt{\frac{\epsilon_{0}}{\epsilon_{p}}}\
  \frac{t_{0}}{t_{p}}\right)^{1/2},
  \label{11}
\end{equation}
where subscripts denote that the corresponding values are given
for the values $\phi(t)$ at the instants of time $t = t_{0}$ and
$t = t_{p}$. Setting $t_{0} \sim 10^{17}$ s (the age of the
Universe), $t_{p} \sim 10^{-44}$ s and using the estimation
$\epsilon _{0}/\epsilon _{p} \sim V_{p}/V_{0}$, where $V_{p} \sim
m_{Pl}^{4}$ and $V_{0} \sim 10^{-5} \, \mbox{GeV}/\mbox{cm}^{3}$
we find that the value of $a(t_{0}) \sim 10^{28}$ cm corresponds
to $a(t_{p}) \sim 10^{-33}$ cm. Thus the dependence of $\epsilon$
on $\phi(t)$, in principle, makes it possible to provide the
missing power (cf. \cite{2,3}) in the $t$ dependence of $a$ and to
solve the problem of size of the Universe.

Taking into consideration the mechanism of quantum tunneling
through the barrier and competing process of the reduction of $V$
with time (which leads to the growth of the barrier $U$ in width
and height) allows to reexamine old and new puzzles of cosmology
from the point of view of quantum theory.

\section{Quantum theory}

\subsection{Quantization and properties of wave function}

In quantum theory, the constraint equation (\ref{51}) comes to be a
constraint on the wave function that describes the universe filled
with a scalar field and radiation \cite{10,11,12}
\begin{equation}
   2\,i\, \partial _{T} \Psi = \left[ \partial _{a}^{2} -
  \frac{2}{a^{2}}\,\partial _{\phi }^{2} - U \right] \Psi \,.
\label{12}
\end{equation}
Here the order parameter is assumed to be zero \cite{2,13,14}.
This equation represents an analog of the Schr\"{o}dinger equation
with a Hamiltonian independent of the time variable $T$. One can
introduce a positive definite scalar product $\langle \Psi | \Psi
\rangle < \infty $ and specify the norm of a state. This makes it
possible to define a Hilbert space of physical states and to
construct quantum mechanics for model of the universe being
considered.

A solution to equation (\ref{12}) can be represented in
the integral form
\begin{equation}
  \Psi (a, \phi , T) = \int_{- \infty}^{\infty }\!dE\,\mbox{e}^{\frac{i}{2} E T}\,
               C(E)\, \psi _{E}(a, \phi ),
\label{13}
\end{equation}
where the function $C(E)$ characterizes the $E$ distribution of
the states of the universe at the instant $T = 0$, while $\psi
_{E}(a, \phi )$ and $E$ are, respectively, the eigenfunctions and
the eigenvalues for the equation
\begin{eqnarray}
 \left( -\,\partial _{a}^{2} + \frac{2}{a^{2}}\,\partial _{\phi }^{2} +
             U - E  \right)  \psi _{E} = 0.
\label{14}
\end{eqnarray}
This equation turns into the famous Wheeler-DeWitt equation for
the minisuperspace model \cite{2,11,13} in special case $E = 0$.

A solution to equation (\ref{14}) can  be represented as
\begin{eqnarray}
    \psi _{E}(a, \phi ) = \int_{- \infty}^{\infty }\! d\epsilon \,
    \varphi _{\epsilon }(a, \phi )\, f_{\epsilon}(\phi ; E),
\label{15}
\end{eqnarray}
where $\varphi _{\epsilon }$ and $\epsilon $ are the eigenfunctions
and the eigenvalues of the equation
\begin{equation}
  \left(-\, \partial _{a}^{2} + U \right) \varphi _{\epsilon } =
  \epsilon \varphi _{\epsilon }.
 \label{151}
\end{equation}
For slow-roll potential $V$, when $\left|d \ln V / d \phi
\right|^{2} \ll 1$, the $\varphi _{\epsilon }$ describes the
universe in the adiabatic approximation and corresponds to
continuum states at a fixed value of the field $\phi $. The
functions $\varphi _{\epsilon }$ can be normalized to the delta
function $\delta (\epsilon  - \epsilon ')$. Their form greatly
depends on the value of $\epsilon $. The quantities
$f_{\epsilon}(\phi ; E)$ can be interpreted as the amplitudes of
the probability that the universe is in the state with a given
values of $\phi$ and $E$ \cite{12}.

Since the potential $U$ has the finite height $U_{max} = 1/4 V$
and finite width then the quantum tunneling through the region
$a_{1} \leq a \leq a_{2}$ of the potential barrier is possible. It
results in that stationary states cannot be realized in the region
$a \leq a_{1}$. If, however, $V(\phi ) \ll 1$, quasistationary
states with lifetimes exceeding the Planck time can exist within
the barrier. The positions  $\epsilon_{n}$ and widths $\Gamma_{n}$
of such states are determined by the solutions to equation
(\ref{151}) for $\varphi_{\epsilon}$ that satisfy the boundary
condition in the form of a wave traveling toward greater values of
$a$. Let us describe these states.

We choose some value $R > a_{2}$. Then
$\varphi_{\epsilon}(a,\phi)$ at fixed $\phi$ can be represented in
the form
\begin{equation}
  \varphi _{\epsilon }(a) = {\cal A}(\epsilon )\,\varphi _{\epsilon
  }^{(0)}(a) \quad \mbox{for} \ 0 < a < R,
\label{16}
\end{equation}
and
\begin{equation}
 \varphi_{\epsilon}(a) =
 \frac{1}{\sqrt{2\pi}}\,\left[\varphi_{\epsilon}^{(-)}(a)-
 {\cal S}(\epsilon)\,\varphi_{\epsilon}^{(+)}(a) \right] \quad \mbox{for}
 \ a > R,
  \label{17}
\end{equation}
where ${\cal A}(\epsilon)$ and ${\cal S}(\epsilon)$ are the
amplitudes depending on $\epsilon$, $\varphi _{\epsilon }^{(0)}$
is the solution that is regular at the point $a = 0$, normalized
to unity, and  weakly dependent on $\epsilon $, while
$\varphi_{\epsilon}^{(-)}(a)$ and $\varphi_{\epsilon}^{(+)}(a)$
describe the wave ``incident'' upon the barrier (the contracting
universe) and the ``outgoing'' wave (the expanding universe)
respectively. Beyond the turning points the WKB approximation is
valid so that one can write
\begin{equation}
  \varphi_{\epsilon}^{(\pm)}(a) = \frac{1}{\sqrt{2}\,(\epsilon -
  U)^{1/4}}\,\exp \left\{\mp\,i \int_{a_{2}}^{a}\! \sqrt{\epsilon -
  U}\,da \pm \frac{i\pi}{4}\right\}.
  \label{18}
\end{equation}
The amplitude ${\cal A}(\epsilon )$ has a pole in the complex
plane of $\epsilon $ at $\epsilon = \epsilon _{n} + i \Gamma_{n}$,
and for $a < R$ the main contribution to the integral (\ref{15})
over the interval $- \infty < \epsilon < U_{max}$ comes from the
values $\epsilon \approx \epsilon _{n}$. The amplitude ${\cal
S}(\epsilon)$ is an analog of the S-matrix \cite{15,16}.

The estimation
\begin{equation}
  \left|\varphi_{\epsilon_{n}}\right|_{a < R} \sim
  \left(\frac{2}{R\,
  \Gamma_{n}}\right)^{1/2}\,\left|\varphi_{\epsilon_{n}}\right|_{a > R}
  \label{19}
\end{equation}
shows that at $\Gamma_{n} \ll 1$ the wave function
$\varphi_{\epsilon}(a)$ has a sharp peak for $\epsilon =
\epsilon_{n}$ and it is concentrated mainly in the region limited
by the barrier. If $\epsilon \neq \epsilon_{n}$ then for the
maximum value of the function $\varphi_{\epsilon}$ we obtain
\begin{equation}
  \left|\varphi_{\epsilon}\right|_{max}^{2} \sim
  \frac{\Gamma_{n}}{R}\,\frac{\sqrt{\epsilon}}{(\epsilon -
  \epsilon_{n})^{2}}\, \left|\varphi_{\epsilon}\right|_{a =
  a_{3}}^{2},
  \label{20}
\end{equation}
where $U(a_{3}) = 0$ and $a_{3} \neq 0$. From this it follows
that for $\Gamma_{n} \ll 1$ the wave function reaches the great
values on the boundary of the barrier, while under the barrier
$\varphi_{\epsilon} \sim O(\Gamma_{n})$.

In the limit of an impenetrable barrier, the function $\varphi
_{n} = \varphi_{\epsilon_{n}}^{(0)}$ reduces to the wave function
of a stationary state with a definite value of $\epsilon _{n}$.
During the time interval $\Delta T < 1/\Gamma_{n}$ the possibility
that the state decays can be disregarded. This corresponds to
defining a quasistationary state as that which takes the place of
a stationary state when the probability of its decay becomes
nonzero \cite{16}.

\subsection{The universe in low-lying quasistationary states}

Calculation of the parameters $\epsilon_{n},\,\Gamma_{n}$ of the
quantum state of the universe can be done by both perturbation
theory by considering the interaction $a^{4}V(\phi)$ as a small
perturbation against $a^{2}$ (in the region $a < a_{1}$ we have
$a^{2}V < 1$) and direct integration of the equation (\ref{151})
 \cite{10,11}. Such calculations show
that the first level with $\epsilon_{0} = 2.62 = 1.31\,\hbar$ and
$\Gamma_{0} = 0.31 = 0.67\,t_{Pl}^{-1}$ emerges at $V = 0.08 = 4.5
\times 10^{-2}\,m_{Pl}^{4}$.

In the early universe, the quantity $V(\phi )$ specifies the
vacuum energy density. The investigations within inflationary
models suggest that the potential $V(\phi(t))$ of the classical
scalar field decreases with time. As the potential $V(\phi )$
decreases, the number of quantum states in the prebarrier region
increases but the decay probability decreases exponentially. The
results of calculations are summarized in the table.

Let us note that the quantum fluctuations of $\phi(t)$ in
exponentially expanding universe can result in that the quantity
$\phi(t)$ and the potential $V \sim \phi^{n}$ will increase
\cite{2,3}. Then the quantum states of the universe in the
prebarrier region cannot form. This case is not interesting for us
and it will not be considered.

\begin{table}
\caption{Parameters $\epsilon_{n}$ and $\Gamma_{n}$ for various
values of the potential $V$. At $V = 5.6 \times
10^{-3}\,m_{Pl}^{4}$ there are six levels in the system, three of
them are displayed.}

\begin{center}

\begin{tabular}{c|c|c|c} \hline
 $V$ $(m_{Pl}^{4})$ & $n$   & $\epsilon_{n}$ $(\hbar)$ & $\Gamma_{n}$ $(t_{Pl}^{-1})$
 \\ \hline
       $4.5 \times 10^{-2}$        &   0    &           1.31           &   $6.7 \times 10^{-1}$
       \\ \hline
       $2.8 \times 10^{-2}$        &   0    &           1.40           &   $1.3 \times 10^{-2}$
       \\ \hline
       $1.7 \times 10^{-2}$        &   0    &           1.45           &   $4.3 \times 10^{-6}$   \\
                                   &   1    &           3.17           &   $2.2 \times 10^{-2}$
       \\ \hline
       $1.1 \times 10^{-2}$        &   0    &           1.47           &  $1.5 \times 10^{-10}$   \\
                                   &   1    &           3.30           &   $2.2 \times 10^{-6}$   \\
                                   &   2    &           4.94           &   $6.5 \times 10^{-3}$
       \\ \hline
       $5.6 \times 10^{-3}$        &   0    &           1.49           &  $2.2 \times 10^{-24}$   \\
                                   &   1    &           3.40           &  $2.2 \times 10^{-19}$   \\
                                   &   2    &           5.26           &  $2.2 \times 10^{-15}$
       \\ \hline
\end{tabular}
\end{center}
\end{table}

The calculations demonstrate that the first instants of the
existence of the universe (counted off from the moment of
formation of first quasistationary state) are especially favorable
for its tunneling through the potential barrier $U$. The emergence
of new levels results in appearance of competition between the
tunneling  processes and transitions between the states. In the
approximation of a slowly varying field $\phi $, transitions in
the system being studied can be considered as those that occur
between the states $| n \rangle $ of an isotropic oscillator with
zero orbital angular momentum which are induced by the interaction
$a^{4} V$. In more strict approach which takes into account the
variations of the $V(\phi)$ the transitions will be carried out at
the expense of the gradient of the potential $V(\phi)$ which
follows from the  quantization of the equation (\ref{51}) taking
into account the evolution of the field $\phi$ in approximation
(\ref{91}) \cite{15}.

Considering the processes of transitions from some initial state
$m(T_{0})$ to final state $n(T)$ (including the case $m = n$) and
tunneling through the barrier from the final state as independent
one can calculate the probability $W_{nm}$ of transition between
the states $m$ and $n$ as:
\begin{equation}
  W_{nm} \thickapprox
  \left|\langle \varphi_{n}|\mathcal{U}_{I}|\varphi_{m}\rangle \right|^{2}\,
  \exp\{-\Gamma_{n}\,\Delta T\},
  \label{26-2}
\end{equation}
where $\Delta T = T - T_{0}$ and $\mathcal{U}_{I}$ is the
evolution operator in the interaction representation \cite{D}. In
the case of two-level system the computation of the total
probability of universe decay, $W_{dec} = 1 - \left( W_{0 0} +
W_{1 0} \right)$, and the probability $W_{1 0}$ demonstrates that
over the time interval $\Delta T \lesssim 50\,t_{Pl}$, the
transitions in the system predominate and only for $\Delta T \sim
10^{2}\,t_{Pl}$ the probability that the universe tunnels through
the barrier becomes commensurate with the probability that it
undergoes the $0 \rightarrow 1$ transition in the prebarrier
region (see figure).

\begin{figure}
\begin{center}
\includegraphics[scale=1.15]{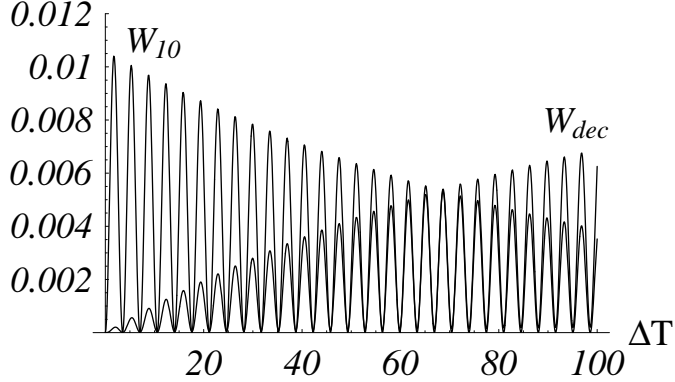}
\end{center}
\vspace{-2.cm} \caption{Probabilities $W_{10}$ and $W_{dec}$
versus the time interval $\Delta T = T - T_{0}$ taken in units of
the Planck time at $V = 1.7 \times 10^{-2}\,m_{Pl}^{4}$.}
\end{figure}

Since the rate at which the level width $\Gamma _{n}$ tends to
zero is greater than the rate at which the $V$ decreases, its
reduction with time results in that transitions become much more
probable than tunnel decays, in which case the former fully
determine the evolution of the quantum universe in the prebarrier
region. If the universe has not tunneled through the barrier
before the potential of the field $\phi $ decreases to a value $V
< 0.01 = 5.6 \times 10^{-3}\,m_{Pl}^{4}$,
a sufficiently large number of levels such that the probabilities
of decays from them can be neglected are formed in the system.
Calculating the amplitudes of transitions over the time interval
$\Delta T $ we find that the $ n \rightarrow n + 1 $ transition is
more probable than the $ n \rightarrow n - 1,\ n + 2$ transitions.
This means that the quantum universe can undergo transitions to
ever higher levels with a nonzero probability. Since the
expectation value of the scale factor $\bar a_{n} = \langle
\varphi _{n} | a | \varphi _{n} \rangle \sim \sqrt{n}$ then it can
be concluded that the characteristic size $\bar a_{n}$ of the
universe that did not undergo a tunnel decay increases as it is
excited to higher levels, i.e. the quantum universe being with
respect to $a$ in classically accessible region before the turning
point $a_{1}$ can evolve so that $\bar a_{n}$ will increase with
time. This can be interpreted as an expansion of the universe. The
probability that, in course of time, the universe will occur in
the region $a > a_{2}$ outside the barrier is negligibly small. In
the limit $V \rightarrow 0$, the universe is completely locked in
the region within the barrier.

The universe in the lowest state with $n = 0$ has a ``proper
dimension'' $d \approx \pi\,\bar{a}_{0} \approx 3 \times 10^{-33}$
cm, the total matter-energy density $\rho \approx
0.65\,m_{Pl}^{4}$, and the problem of the initial cosmological
singularity does not emerge. The classical turning points are
$a_{1} \simeq 1.4 \times 10^{-33}$ cm and $a_{2} \simeq 2.2 \times
10^{-33}$ cm. The value of $a_{1}$ determines the maximal
dimension of the universe occurring in the lowest state in
prebarrier region, while $a_{2}$ characterizes its initial
dimension after tunneling from this state. If a quantum universe
tunnels from the states with $n
> 0$, the dimensions of the region from which tunneling occurs can
considerably exceed the Planck length. The constants $E$ and $V$
appearing in the Einstein equations will be determined by the
corresponding quantum stage.

Thus it turned out that quantum universe originally filled with a
radiation and a matter in form of a scalar field with a potential
$V(\phi(t))$ decreasing with time has nonzero probability to
evolve remaining in prebarrier region. The expansion here is
ensured by the transitions from lower states to higher states by
means of interaction between gravitational and scalar fields. The
system can be found in the highly excited states with $n \gg 1$ as
a result of such evolution. In states with $n \gg 1$ the potential
$V \ll 1$ and $\Gamma_{n} \sim 0$. The state of the universe will
be characterized by the quantum number $n$ which determines its
geometrical properties and new quantum number $s$ responsible for
the state of matter.

\subsection{Highly excited states}

The potential of the $\phi$ will be chosen in the form $V(\phi) =
(m^{2}/2)\,\phi^{2}$. From the condition
 $V \ll 1$, it follows that the mass of the field must be constrained
by the condition $m \ll m_{Pl}^{2}/|\phi |$. In this case the
states of the matter are determined by the solutions of the
Schr\"{o}dinger equation for a harmonic oscillator at given value
of $n$ \cite{12}. It describes the oscillations of the $\phi$ near
the minimum of the potential $V(\phi)$. This process can be
interpreted as the production of particles. At $E = 0$, a similar
mechanism leads to the production of particles by the inflaton
field, which is identified with the scalar field $\phi $ \cite{2}.
Assuming as before that the $V(\phi)$ is slow-roll potential we
find the condition of quantization of $E$
\begin{equation}
    E = 2 N - (2 N)^{1/2} (2 s + 1) m,
\label{21}
\end{equation}
where $N = 2n + 1$, and the values of the quantum number $s$ are
restricted by inequality $s + 1/2 \ll \sqrt{2 N^{3}}/m$ which
reflects the fact that the mass $m$ of the produced particles is
finite. For small $s$ the equality $E \approx 2N$ holds to a high
precision, so that the universe is dominated by radiation. A
transition from the radiation-dominated universe to the universe
where matter (in the form of particles produced by the field $\phi
$) prevails occurs when the second term in (\ref{21}) becomes
commensurate with the first one. The physical interpretation of
the condition (\ref{21}) will be considered below.

For the universe with given quantum numbers $n \gg 1$ and $s \gg
1$ the wave function $\psi_{E}$ has the form \cite{12}
\begin{equation}
  \psi_{E}(a,\phi) = \varphi_{n}(a)\,f_{ns}(\phi),
  \label{27-2}
\end{equation}
where
\begin{equation*}
  \varphi_{n}(a) = \left(\frac{2}{N}\right)^{1/4}\,\cos \left(\sqrt{2
  N}\,a - \frac{N \pi}{2}\right),
\end{equation*}
\begin{equation*}
  f_{ns}(\phi) = \left(\frac{m (2 N)^{3/2}}{2
  (2s + 1)}\right)^{1/4}\,\cos \left(\sqrt{2s + 1}\,(2 m^{2}
  N^{3})^{1/4}\,\phi- \frac{s \pi}{2}\right).
\end{equation*}
This wave function is normalized to unity with allowance for the
fact that the probability of finding the universe in the region $a
> a_{2}$ is negligibly small.

The condition (\ref{21}) can be rewritten in the terms of
``observable'' quantities: the cosmic scale factor  $\langle a
\rangle = \sqrt{N/2}$, where averaging was performed over the
state (\ref{27-2}), and the total mass of the matter $M = m(s +
1/2)$,
\begin{equation}
   E = 4 \langle a \rangle \left[ \langle a \rangle - M \right].
\label{22}
\end{equation}
The classical universe is characterized by the total energy
density $\rho = T_{0}^{0} + \tilde T_{0}^{0}$. Replacing all
quantities by corresponding operators for the quantum universe we
set $\rho_{tot} = \langle \rho \rangle$. It gives $\rho_{tot} =
\rho_{sub} + \rho_{rad}$, where
\begin{equation}
    \rho _{sub} = \frac{193}{12}\,\frac{M}{\langle a \rangle
    ^{3}}\  \quad \mbox{and} \,\quad
         \rho_{rad} = \frac{E}{\langle a \rangle ^{4}}\,.
\label{23}
\end{equation}
Here in accordance with the Ehrenfest theorem we assume that
expectation value $\langle a \rangle$ follows the laws of
classical theory and the expectation values of the functions
$T^{0}_{0}$ and $\tilde{T}^{0}_{0}$ of $a$ can be replaced by the
functions of $\langle a \rangle$.

In the case, when $\rho_{sub} \gg \rho_{rad}$ we
have $\langle a \rangle = M$. This relation holds to a high
precision $\sim 10^{-5}$ in the observed part of our Universe,
where $\langle a \rangle \sim 10^{61} \sim 10^{28}$ cm and
 $M \sim
10^{61} \sim 10^{56}$ g. The quantum numbers of such universe are the following:
 $n \sim \langle a \rangle ^{2} \sim 10^{122}$ and $s \sim \langle a
\rangle / m \sim 10^{80}$ taking the proton mass for $m$. The
value of $n$ agrees with  existing estimates for our Universe,
while $s$ is equal to equivalent number of baryons \cite{13,18}.

Thus for the matter-dominant era we have the following relation between
$\langle a \rangle$ and $\rho_{sub}$:
\begin{equation}
   \langle a \rangle = \left(\frac{193}{12}\,\frac{1}{\rho _{sub}}
         \right)^{1/2}.
\label{231}
\end{equation}
This relation can be reduced to the standard form \cite{18,19}
\begin{equation}
  \frac{1}{\langle a \rangle^{2}} = \left(\Omega_{0} - 1\right)\,H^{2},
\label{232}
\end{equation}
where $H^{2} = \rho_{c}$ is the critical density and $\Omega_{0} =
\rho_{sub}/\rho_{c}$ is the density parameter equal to $\Omega_{0}
= 1.066$. That is, the geometry of the universe with $n \gg 1$ and
$s \gg 1$ is close to Euclidean geometry (a flat universe).

If one neglects the contribution from the kinetic term of the
scalar field ($\pi_{\phi}^{2} = 0$) then the corresponding $\Omega
_{0} \simeq 0.077$. This value exceeds the contribution from the
luminous matter (stars and associated material) \cite{4} and it is
close to the value for minimum amount of dark matter required to
explain the flat rotation curves of spiral galaxies.  Although the
potential $V(\phi )$ undergoes only small variations in response
to changes in the field $\phi $, the field $\phi $ itself changes
fast, oscillating about the point $\phi  = 0$, so that the
approximation in which $\pi _{\phi }^{2} = 0$ is invalid. The
application of the present model in this approximation would
result in the radiation-dominated universe; that is, it would not
feature a mechanism capable of filling it with matter after a slow
descent of the potential $V(\phi )$ to the equilibrium position,
which corresponds to the true vacuum.

It is interesting to find the physical interpretation on
(\ref{22}). Passing on to the ordinary physical units we rewrite
it in the form
\begin{equation}
  a = G\,M_{tot},
  \label{30}
\end{equation}
where $M_{tot} = M + U_{rad}$, $U_{rad} \equiv E /2 a$. It is easy
to see that (\ref{30}) is the condition of the equality between
proper gravitational energy of the thin spherical layer (with the
total mass $M_{tot}$) on the sphere with radius $a$ and the sum of
energies of particles $M$ and energy of radiation $U_{rad}$. In
the modern era $M \sim 10^{80} \, \mbox{GeV} \gg U_{rad} \sim
10^{75}$ GeV. If we extrapolate (\ref{30}) on Planck era and set
$a = l_{Pl}$ then $M_{tot} = m_{Pl}$. For the lowest quantum state
we find that $U_{rad} \approx m_{Pl}/2$. Since $s = 0$ the
parameter $m \approx m_{Pl}$. It means that vacuum energy of the
scalar field and the energy of radiation make the comparable
contribution to the total energy of universe with $n = 0$. In this
state the scalar field $|\phi| \approx 0.3\,m_{Pl}$.

\subsection{The nonhomogeneities of matter density}

The approach developed here makes it possible to obtain realistic
estimates for the proper dimensions of the nonhomogeneities of
matter density, for the amplitude of the fluctuations of the
cosmic background radiation temperature and points to a new
possible mechanism of their origin, namely by means of finite
values of the widths of quasistationary states. For a small, but
finite value of the width $\Gamma $ the quasistationary state does
not possess a definite value of $\epsilon $. The corresponding
uncertainty $\delta \epsilon $ can serve as source of fluctuations
of the metric $\delta a$ \cite{12}. By associating $\epsilon +
\delta \epsilon$ with the scale factor $a + \delta a$ and by using
the solution (\ref{92}) for $a(0) = 0$ we find the amplitude of
fluctuations of the scale factor in the form
\begin{equation}
   \frac{\delta a}{a} = \frac{1}{4}\,
    \frac{\delta \epsilon / \epsilon}{1 - \tanh (\sqrt{V}t) /2
    \sqrt{V \epsilon }}.
  \label{24}
\end{equation}
Since $\delta \epsilon \lesssim \Gamma $, the fluctuations $\delta
a$ that were generated at the early stage of the evolution of the
Universe will take the greatest values. For the lowest
quasistationary state with $\epsilon = 2.62$, $\delta \epsilon
\lesssim 0.31$, $V = 0.08$, at $t \sim 1$ we obtain
\begin{equation}
  \frac{\delta a}{a} \lesssim 0.04.
  \label{25}
\end{equation}
Since the dimension of large-scale fluctuations changed in direct
proportion $a$, this relation has remained valid up to the present
time. For the current value of $a \sim 10^{28}$ cm we find that
$\delta a \lesssim 130$ Mpc. On the order of magnitude, the above
value corresponds to the scale of superclusters of galaxies.
Smaller values of $\delta \epsilon $ are peculiar to quantum
states with smaller $V$. The fluctuations $\delta a $
corresponding to them are smaller than (\ref{25}) and are expected
to manifest themselves against the background of the large-scale
structure. They can be associated with clusters of galaxies,
galaxies themselves, and clusters of stars.

\subsection{Fluctuations of the cosmic background radiation temperature}

The energy density of radiation can be expressed as
\begin{equation}
   \rho_{rad} = \frac{4 \pi ^{4}}{30}\,g_{*}\,\mbox{T}^{4},
  \label{251}
\end{equation}
where $\mbox{T}$ is temperature and $g_{*}$ counts the total number
of effectively massless degrees of freedom
\cite{1,2,4}. Using the relation (\ref{23}) for $\rho_{rad}$ we obtain
\begin{equation}
   E = \frac{4 \pi ^{4}}{30}\,g_{*}\,\left(a\,\mbox{T}\right)^{4},
  \label{252}
\end{equation}
where we omit the brackets for simplicity. Leaving the main terms we can write
\begin{equation}
   \frac{\delta \mbox{T}}{\mbox{T}} \simeq
   \frac{1}{4}\,\frac{\delta \epsilon}{\epsilon } -
     \frac{\delta a}{a}\ .
  \label{253}
\end{equation}

For $\sqrt{V}\,t \ll 1$ it follows the estimation
\begin{equation}
   \frac{\delta \mbox{T}}{\mbox{T}} \simeq  \frac{t}{2 \sqrt{\epsilon }}\,
     \frac{\delta a}{a}\ .
  \label{26}
\end{equation}
For the time $t \sim 10^{5}$ yr corresponding to the instant of
recombination of primary plasma (separation of radiation from
matter), and for the observed value of $\epsilon = 2.6 \times
10^{117}\,\hbar$, for (\ref{25}) we find
\begin{equation}
   \frac{\delta \mbox{T}}{\mbox{T}} \lesssim  2.8 \times 10^{- 5}.
  \label{27}
\end{equation}
Here $\sqrt{V}\,t \sim 0.7 \times 10^{-3}$.

Upon recombination, the fluctuations of the temperature undergo no
changes; therefore, measurement of the quantity $\delta \mbox{T} /
\mbox{T}$ for the present era furnishes information about the
Universe at the instant of last interaction of radiation with
matter. The estimate in (\ref{27}) is in good agreement with
experimental data from which the trivial dipole term $\sim
10^{-3}$ caused by the solar system motion was subtracted
\cite{20}.

\subsection{Entropy}

The total entropy $S$ per  comoving volume $2 \pi^{2} a^{3}$
\cite{1,2,4} can be expressed as
\begin{equation}
   S = \frac{4 \pi ^{4}}{45}\,g_{*s}\,\left(a\,\mbox{T}\right)^{3},
  \label{271}
\end{equation}
where $g_{*s}$ can be replaced by $g_{*}$ for the most of the
history of the Universe when all particles species had a common
temperature.

From (\ref{252}) and (\ref{271}) it follows  a simple relation
between the $E$ and the total entropy $S$
\begin{equation}
  \frac{E}{S} = \frac{3}{2}\, \frac{g_{*}}{g_{*s}}\,a \mbox{T}.
  \label{28}
\end{equation}
For the adiabatic expansion, $a \mbox{T} = \mbox{const}$, the
ratio $E/S$ is conserved. Excluding $a \mbox{T}$ from (\ref{28})
we obtain the relation,
\begin{equation}
  S^{4} = \left(\frac{2}{3}\right)^{5} \frac{\pi^{4}}{5}\,g_{*s}
  \left(\frac{g_{*s}}{g_{*}}\right)^{3}  E^{3}.
  \label{29}
\end{equation}
In the era with $\mbox{T} \sim 10^{19}$ GeV we have $S \sim 1$,
but at present time for $\mbox{T} \sim 10^{-13}$ GeV the entropy
$S \sim 10^{88}$. The large value of $E$ today explains the large
value of entropy of the Universe.

\subsection{Acceleration or deceleration?}

Recent measurements \cite{P,R} indicate that today the Universe is
accelerating. Let us note that another possible explanation of the
observed dimming of the type Ia supernovae at redshifts $z \sim
0.5$ is an unexpected supernova luminosity evolution \cite{RFLS}.
At the present time  the first interpretation of observed
phenomenon is considered as more preferable.

In terms of classical cosmology the accelerated expansion is
described by the negative values of the deceleration parameter
\begin{equation}
  q = - \frac{1}{H^{2}}\,\frac{\partial_{t}^{2} a}{a}\ .
  \label{42}
\end{equation}
In order to agree the experimental data with the theory it was
proposed the concept of the dark energy which is nearly smoothly
distributed in space. This dark energy component must have
negative pressure that overcomes the gravitational self-attraction
of matter and causes the accelerated expansion of the Universe. It
is commonly assumed that the vacuum energy density in the form of
a non-zero cosmological constant or due to a slow-roll scalar
field called "quintessence" can be responsible for the dark energy
\cite{O,C,BB}.

Let us examine this problem from the point of view of the approach
developed in this paper. To this end we rewrite (\ref{42}) in the
form
\begin{equation}
  q = 1 + a\,\frac{\partial_{T} \pi_{a}}{\pi_{a}^{2}}.
  \label{43}
\end{equation}
Bearing in mind canonical equation for $\partial_{T} \pi_{a}$ from
(\ref{52}) and having differentiated the equation (\ref{14}) with
respect to $a$, we obtain that the derivative $-\,
\partial_{T} \pi_{a}$ must be substituted by the
quantum mechanical operator
\begin{equation}
  \Pi \equiv \frac{1}{2} \left(-\partial_{a}^{2} + \frac{2}{a^{2}}\,
  \partial_{\phi}^{2} + a^{2} - a^{4}V - E\right) \partial_{a}.
  \label{44}
\end{equation}
Then according to quantum mechanical principles the quantum analog
of deceleration parameter can be calculated as
\begin{equation}
  \langle q \rangle = 1 - \frac{\langle a\, \Pi \rangle}{\langle \pi_{a}^{2}
  \rangle},
  \label{45}
\end{equation}
where averaging is performed over states $\psi_{E}$ and it is
assumed that offdiagonal matrix elements from $q$ vanish. (This
corresponds to the representation of deceleration parameter as a
scalar quantity.)

In state with large quantum numbers $n \gg 1$ and $s \gg 1$, for
the matter-dominant universe, where $E / 4 \langle a \rangle ^{2}
\ll 1$, using the wave function (\ref{27-2}) we obtain
\begin{eqnarray}
  \langle q \rangle =  1 - \frac{1}{2}
   \left[\cos\left(2 \pi  \langle a \rangle
  ^{2}\right) + \frac{2}{3} \cos\left(\left(2 \pi - 8\right) \langle a \rangle
  ^{2}\right) \right].
  \label{46}
\end{eqnarray}
The expression in square brackets in (\ref{46}) which contains two
cosines rapidly oscillates with small period $\sim 2 l_{Pl}$.
Averaging (\ref{46}) over small interval near some fixed value of
$\langle a \rangle ^{2}$ we have
\begin{equation}
  \overline{\langle q \rangle} = 1.
  \label{51-1}
\end{equation}
This value can be associated with the deceleration parameter in
classical theory. It agrees with the classical conceptions of
general relativity about the expansion rate of the Universe in
matter-dominated era with zero cosmological constant \cite{18,19}.

The quantity (\ref{51-1}) does not take into account the quantum
fluctuations of the scale factor
\begin{equation}
  \Delta a = \sqrt{\langle a^{2} \rangle - \langle a \rangle ^{2}}
  \label{52-1}
\end{equation}
that specify root-mean-square deflection of the distribution
$|\psi _{E}(a, \phi)|^{2}$ as function of $a$. In this case
$\psi_{E}$ represents the wave packet which describes the universe
being localized in space $a$ near the expectation value $\langle a
\rangle$ with deflection $\Delta a$. We shall show that at certain
conditions (parameters of the universe) the fluctuations $\Delta
a$ can affect essentially the character of the expansion of the
universe. It can provide in particular the accelerated expansion
observed nowadays \cite{P,R}.

We shall denote the scale factor taking into account the
fluctuations $\Delta a$ as $\tilde{a} = \tilde{a}(t)$, while the
fluctuations will be associated with quantity $\Delta a =
\tilde{a} - a$, where $a = a(t)$ is the scale factor without
regard for fluctuations. Fixing some instant $t_{0}$ for small
intervals $\Delta t = t - t_{0}$ we can write the expansion
\cite{18}
\begin{equation}
  a = a_{0}\left[1 + H_{0}\,\Delta t - \frac{1}{2}\,q_{0}\, H_{0}^{2}\, \Delta
  t^{2} + \ldots \right],
  \label{53}
\end{equation}
and similarly
\begin{equation}
  \tilde{a} = \tilde{a}_{0}\left[1 + \tilde{H}_{0}\,\Delta t -
  \frac{1}{2}\,\tilde{q}_{0}\,\tilde{H}_{0}^{2}\, \Delta
  t^{2} + \ldots \right],
  \label{54}
\end{equation}
where the Hubble constant $\tilde{H}_{0}$ and the deceleration
parameter $\tilde{q}_{0}$ are calculated with respect to scale
factor $\tilde{a}$ and subscript $0$ means that corresponding
values are taken at $t = t_{0}$. From (\ref{53}) and (\ref{54}) it
follows that
\begin{equation}
  \Delta a = \Delta a_{0} + \left(\partial_{t} \Delta a\right)_{0}
  \Delta t + \frac{1}{2}\,\left(\partial_{t}^{2} \Delta a\right)_{0}
  \Delta t^{2} + \ldots,
  \label{55}
\end{equation}
where
\begin{equation}
  \left(\partial_{t} \Delta a\right)_{0} =
  \tilde{a}_{0} \tilde{H}_{0} - a_{0} H_{0}, \quad
  \left(\partial_{t}^{2} \Delta a\right)_{0} =  a_{0}\, q_{0}\,
  H_{0}^{2}
  - \tilde{a}_{0}\, \tilde{q}_{0}\, \tilde{H}_{0}^{2}.
  \label{56}
\end{equation}
Let us suppose that the fluctuations $\Delta a $ do not affect the
rate of expansion, that is $\tilde{a}_{0} \tilde{H}_{0} = a_{0}
H_{0} = (\partial_{t} a)_{0} = \mbox{const}$. In addition it is
natural to assume that the Hubble constant also does not depend on
these fluctuations, i.e. $\tilde{H}_{0} = H_{0}$. These
assumptions are based on astrophysical observations which do not
record the necessity to modify the classical conception of the
Hubble's law. With regard to these two facts we obtain
\begin{equation}
  \left(\partial _{t}^{2} \Delta a \right)_{0} = 2\,
   \frac{\Delta a}{\Delta t^{2}}\, .
  \label{57}
\end{equation}
Then according to (\ref{56}) the deceleration parameter
$\tilde{q}_{0}$ which takes into account the quantum fluctuations
$\Delta a$  will be determined by the expression
\begin{equation}
  \tilde{q}_{0} = q_{0} - \frac{2}{(H_{0} \Delta t)^{2}}\,
  \frac{\Delta a}{a_{0}}\,.
  \label{58}
\end{equation}
Using the wave function (\ref{27-2}) from (\ref{52-1}) we find
that for the state of the universe with large quantum numbers $n$
and $s$ the fluctuations
\begin{equation}
  \Delta a = \frac{\langle a \rangle}{\sqrt{3}}
  \label{59}
\end{equation}
and the deceleration parameter
\begin{equation}
   \tilde{q}_{0} = q_{0} - \frac{2}{\sqrt{3}}\, \frac{1}{(H_{0} \Delta t)^{2}}\,
  \frac{\langle a \rangle}{a_{0}}\,.
  \label{60}
\end{equation}
For numerical estimation in the capacity of $\Delta t$ one can
take the age of the Universe and the mean $\langle a \rangle$ may
be put to be equal to classical value $a_{0}$. Then for modern
value $\Delta t = 14\ \mbox{Gyr} = 1.78 \times 10^{61}$ \cite{4}
and $H_{0} = h_{0} \times (9.778\ \mbox{Gyr})^{-1} = 0.8\, h_{0}
\times 10^{-61}$ we obtain
\begin{equation}
  \tilde{q}_{0} = q_{0} - \frac{0.81}{h_{0}^{2}}\,.
  \label{61}
\end{equation}
For the conventional value $H_{0} =
65\,\mbox{km}\,\mbox{s}^{-1}\,\mbox{Mpc}^{-1}$ \cite{F,M} it gives
\begin{equation}
  \tilde{q}_{0} = q_{0} - 1.92.
  \label{62}
\end{equation}
The parameter $q_{0}$ corresponds to the case when the
fluctuations $\Delta a = 0$ and according to (\ref{51-1}) it
equals to $q_{0} = \overline{\langle q \rangle} = 1$. As a result
we find
\begin{equation}
  \tilde{q}_{0} = -0.92.
  \label{63}
\end{equation}
This value takes into account the presence of quantum fluctuations
of metric and it is in good agreement with SNe Ia observations
\cite{P,R}.

Thus, the observed accelerated expansion can be explained without
implication of any additional concepts about matter-energy
structure of the Universe considering this acceleration as
macroscopic manifestation of its quantum nature.

The represented calculations relate to the universe with large
quantum numbers. In preceding epoch the cosmological parameters
$H_{0}$, $\Delta t$ and $\Delta a$ took different values. If one
assumes that up to coefficients on the order of $\sim 1$ the
relation $(H_{0}\,\Delta t)^{2} \approx h_{0}^{2}$ holds for
earlier instants of time and the fluctuations $\Delta a$ as before
have been on the order of $\sim \langle a \rangle / 2$ then from
(\ref{58}) it follows that in epoch with the Hubble constant
$H_{0} \gtrsim 100\,\mbox{km}\,\mbox{s}^{-1}\,\mbox{Mpc}^{-1}$ the
universe have to be decelerated. For more accurate calculation of
the deceleration parameter of the universe in such states the
averaging in (\ref{52-1}) must be performed over the wave
functions $\psi_{E}$ which take into account that the variables
$a$ and $\phi$ in the equation (\ref{14}) are not separated in
general.

\section{Concluding remarks}

The main constructive element of our model which allows to avoid
most of cosmological problems is an idea that $E$ increased during
the evolution of the Universe. The quantity $E$ determines the
energy-momentum tensor of radiation and can be found as an eigenvalue
for the equation (\ref{14}). The above numerical estimations of
the parameters of the quantum universe filled with the radiation
and scalar field show that the averaged massive scalar field used
instead of the aggregate of real physical fields mainly correctly
describes global characteristics of our Universe. It effectively
includes visible baryon matter and dark matter. The kinetic energy
term of the scalar field provides the modern value of the total
energy density of the universe which is very close to the critical
value. The status of the field $\phi $ changes as we go over from
one stage of universe evolution to another. In the early universe,
the field $\phi $ ensures a nonzero value of the vacuum-energy
density due to $V(\phi )$ values at which the equation (\ref{151})
for $\varphi _{\epsilon }(a, \phi )$ admits nontrivial solutions
in the form of quasistationary states. In a later era, when the
field $\phi $ descends to a minimum of the potential $V(\phi )$
and begins to oscillate about this minimum, it appears to be a
source of the particles of some averaged matter filling the
visible volume of the universe, which has linear dimensions on the
order of $\sim \langle a \rangle $. The galaxies, their clusters,
and other structures in the Universe are subject to quantum
fluctuations (due to the finite widths of the quasistationary
states) that have grown considerably.

The quantum fluctuations which specify the spread of the wave
function of the universe in space of scale factor can ensure the
accelerated expansion of the universe. In this sense they manifest
themselves similar to dark energy. The theory gives the value of
the deceleration parameter $q = 1$ (the universe is slowing down)
for essentially classical cosmological macrosystem and predicts $q
\approx - 1$ (the universe is speeding up) explaining the
accelerated expansion as macroscopic manifestation of quantum
nature of the universe.\\

\begin{center}
\textbf{\large Acknowledgements}
\end{center}

We should like to express our gratitude to Alexander von Humboldt
Foundation (Bonn, Germany) for the assistance during the research.

\end{document}